\documentclass[12pt]{article}
\usepackage{amsfonts,graphics}
\newcommand{\eps}{{\cal{E}}}

\newcommand{\calp}{{\cal{P}}}
\newcommand{\cald}{{\cal{D}}}
\newcommand{\calw}{{\cal{W}}}
\newcommand{\call}{{\cal{L}}}

\begin{document}


\pagestyle{empty} \setcounter{page}{0}

\null


\vspace{50mm}

\begin{center}

{\Large \textbf{The Wess Zumino consistency  \\[5mm]
condition: a paradigm in \\[5mm]
renormalized perturbation
theory\footnote{Talk given at Symposium in Honor of Julius Wess on
the Occasion of his 70th Birthday, 10-11 January 2005 at Max
Planck Institute for Physics (Werner Heisenberg Institut)
Fohringer Ring 6 - D-80805 MUENCHEN.}}}

\vspace{10mm}

{\large R. Stora$^{a,b}$}

\vspace{10mm}

\emph{$^a$ Laboratoire d'Annecy-le-Vieux de Physique Th{\'e}orique
(LAPTH)}

\emph{CNRS, UMR 5108, associ{\'e}e {\`a} l'Universit{\'e} de Savoie}

\emph{9, chemin de Bellevue, BP 110, F-74941 Annecy-le-Vieux
Cedex, France}

\vspace{7mm}

\emph{$^b$ CERN, Theory Division}

\emph{1211 Geneva 23, Switzerland}

\end{center}

\vspace{20mm}


\vspace{10mm}

\emph{}


\vfill \vfill


\rightline{June 2005}

\vfill

 \clearpage \pagestyle{plain} \baselineskip=18pt

\section{Introduction}

There is a jewel drowned within Julius' abundant scientific
contribution:

\begin{center}
"Consequences of Anomalous Ward identities"\\
by\\
J. Wess (CERN); B. Zumino (CERN)\\
Phys. Lett. {\bf B37} (1971) 95 \cite{1}
\end{center}
shortly followed by L.C. Biedenharn's\cite{2} (University of
Karlsruhe), talk at the Colloquium on group theoretical methods in
Physics, Centre de Physique Th\'eorique du CNRS, Marseille, 5-9
June 1972, whose author acknowledges discussions with Julius Wess
and quotes Julius's lectures\cite{3} on Anomalous Ward identities
at the International Universit\"at f\"ur Kern Physik, Schladming,
21 Feb.-4 March 1972.

This contains two parts:

a) the observation that current algebra anomalies, when occurring
as right hand sides of Ward identities associated with a Lie
algebra -in that case a gauge Lie algebra with structure group
$(SU3)_L \times (SU3)_R$-, are constrained by "Wess Zumino
consistency relations" which reflect the algebraic structure of
the Lie algebra, in combination with the locality of
(perturbative) field theory. This can be used to check the
consistency of particular Feynman graph calculations.

L.C. Biedenharn's remark\cite{2} was that these consistency
relations were nothing else than Lie algebra cocycle conditions
for the gauge Lie algebra under study.

b) the construction of a compensating action involving an element
of the gauge group.

Part b) had a glamorous carreer thanks to E. Witten's\cite{4}
observation in 1984 of its ties with the topological properties of
the gauge group, (besides its applications to physics already
mentioned in the Letter). However interesting this is, we shall
insist on the central role played by part a) within the known
formulation of renormalized perturbation theory mathematically
founded on causality-(locality)\cite{5}.

In section \ref{section2} we shall recall some of the main
ingredients which go into the construction of renormalized
perturbation expansions.

In section \ref{section3}, we shall recall some of the standard or
less standard examples which illustrate the importance of the WZCC
(Wess Zumino Consistency Conditions).

Some further remarks are collected in the concluding section
\ref{section4}.

\section{Renormalized perturbation theory: locality and power counting\label{section2}}

Renormalized Perturbation Theory (R.P.T.) exists in essentially
two distinct forms, both based on causality (locality) and power
counting, the latter limiting the ambiguities which result from
the singularity of local interactions.

In the first, more natural, set up \cite{6}\cite{7}, the building
blocks are operators acting in some prescribed Fock space in
correspondence with some quantized free fields. These operators
are labelled by Wick polynomials of these free fields. The causal
properties of the latter together with such general properties as
translation covariance, positivity of the energy, relativistic
covariance, result into a strict characterization of the
ambiguities intrinsically attached to the definition of these
operators, ("time ordered products" of Wick polynomials of the
free fields) due to the distribution character which follows form
the general principles\cite{11}.

In the second "functional set up\cite{6}\cite{8}, "time ordered
products" are defined as functionals of some correspondingly
labelled classical fields, and labelled by classical local
polynomials of the latter. Necessary ambiguities are shown to be
choosable within a restricted set, also characterized by locality
and power counting although these limitations are not consequences
of locality without some further assumptions\cite{8}.

The first, "on shell", (dispersive) point of view as well as the
second, off shell, (Feynman) point of view are both treated in the
classical article by H. Epstein, V. Glaser\cite{6}, a landmark in
the long history of the role of causality in the construction of
the renormalized perturbative series. Another long period\cite{12}
has finally given rise to a reexamination of these constructions,
exemplified by two extensive studies, by M. D\"utsch, F.
Boas\cite{7} and M. D\"utsch, K. Fredenhagen\cite{8},
respectively, from which one can reconstruct the previous history.

This has a direct bearing on the definition of the physical
content of these constructions\cite{13} -notwithstanding all the
limitations attached to the perturbative approach -since "physics"
should be devoid of ambiguities.

Although less strictly founded, the off shell formalism\cite{8} is
in a much better shape that the more physical on shell
formalism\cite{7}. Thanks to the introduction of two further
assumptions\cite{8}

a) Time ordered products are multilinear functions of their
arguments.

b) Time ordered products are restricted by the Action Ward
identity (see section \ref{section3}), one can organize the
ambiguities within a renormalization group structure\cite{14},
which yields natural definitions of models and their physical
contents.

The connection between the off shell and the on shell formalism is
much less advanced.

To be brief, imposing the multilinearity of the operator time
ordered products in their sets of Wick ordered local arguments,
and the Hamiltonian Ward identity, the natural analog of the
Action Ward identity (see section \ref{section32}), is not proved
at the moment compatible with Lorentz covariance, which is easy in
the off shell formalism. Seen from an alternative point of view, a
manifestly covariant set of on shell Ward identities has not yet
been derived from the off shell formalism, which would ascertain
the existence of an on shell renormalization group summarizing the
known on shell ambiguities \cite{14}.

In spite of this incomplete knowledge, the role of Ward identities
in the definition of the physical content of these constructions
immediately comes to the forefront as their fulfillment results
into a reduction of the renormalization group. In many interesting
cases the possibility to fulfill a set of Ward identities proceeds
directly through the construction of properly constrained sets of
time ordered products via for instance the use of an intelligent
regularization -renormalization procedure. Otherwise, the Wess
Zumino consistency conditions (WZCC) are the only known general
tool able to detect and or eliminate potential anomalies. This is
tantamount  to saying that the Wess Zumino consistency conditions
are central in the definition of all models which are physically
admissible in view of some anomaly cancellation conditions,
namely, at the moment most of the physically interesting
models\cite{10}.

Whereas the Wess Zumino consistency conditions (WZCC) appeared in
the context of the description of the quotient of field space by
the action of a Lie algebra -a gauge Lie algebra, in that case-,
more general quotients have appeared since, -and will appear-
giving rise to more general differential algebras which have been
contemplated in the mathematical literature at great length since
the fifties.

In section \ref{section3}, we shall review a sample zoo of such
structures mostly but not only originating from symmetries, which
gives a modest measure of the impact of Phys. Lett. {\bf B37}
(1971) 95\cite{1}.

\section{Examples \label{section3}}

\subsection{$SL2 \mathbb{C}$ covariance (Group cohomology)\label{section31}}

In conventional methods appealing to regularizations, it may
happen that one needs to break Lorentz covariance (e.g. in the
momentum space version of BPHZ\cite{9}). Then, because $H^1(SL2
\mathbb{C},$ $F) =0$ for every finite dimensional representation
space $F$ of $SL2\mathbb{C}$, one can always recover covariance by
the inclusion of suitable local counter-terms.

\subsection{The Action Ward Identity\cite{8}\label{section32}}

In the off shell version where $T$ products are labelled by
$n$-uples of local classical polynomials $\calp_i(\varphi)$ of
fields $\varphi$ and their derivatives, with values functionals of
background field $\Phi$, one wishes to impose
\begin{equation}   \label{eq1}
\partial^{(1)}_\mu \ T \ \calp_1(\varphi)(x_1) \ldots
\calp_n(\varphi)(x_n) = T \ \partial^{(1)}_\mu \ \calp_1(\varphi)
(x_1) \ldots \calp_n(\varphi)(x_n)
\end{equation}

The proof that it is possible is given by M. D\"utsch, K.
Fredenhagen\cite{8}.

In the operator version where operator $\widehat{T}$ products are
operators in Fock space labelled by local Wick polynomials of the
free fields $\widehat{\varphi}$, one can similarly impose the
Hamiltonian Ward identity
\begin{equation}\label{eq2}
\overrightarrow{\partial}^{(1)} \ \widehat{T} :
\calp_1(\widehat{\varphi}) : (x_1) \ldots : \calp_n
(\widehat{\varphi}) : (x_n) = \widehat{T}
:\overrightarrow{\partial}^{(1)} \ \calp_1 (\widehat{\varphi}) :
(x_1) \ldots \calp_n (\widehat{\varphi}) : (x_n)
\end{equation}
where $\overrightarrow{\partial}$ denotes space derivatives. (For
the on shell set up, see M. D\"utsch, F.M. Boas\cite{7}).

One could alternatively impose $SL2 \mathbb{C}$ covariance
(according to example \ref{section31}), but one does not know at
the moment whether this is compatible with the Hamiltonian Ward
identity and allows for the absorption of the permissible
ambiguities into an on shell renormalization group.

These identities have different equivalent meanings:

a) the scattering operator only depends on the action -resp.
Hamiltonian-, not on the Lagrangian -resp. Hamiltonian-density.

b) (Energy) momentum conservation holds at each vertex of a
renormalized Feynman graph.

This allows for the absorption of ambiguities into a
renormalization group whose structure is based on the composition
of formal power series (modulo the completion of our knowledge on
the on shell formalism) :

\subsection{Mass shell restriction \label{section33}}

In the off shell formalism, one can restrict the ambiguities in
such a way that $T$ products with at least one argument in the
ideal generated by the free fields equations of motion\cite{7}
belong to the corresponding ideal in the space of functionals. But
this is in clash with the Action Ward identity.

It is a strongly supported but not yet proved conjecture that, in
the on shell formalism, there is an $SL2 \mathbb{C}$ covariant set
of Hamiltonian Ward identities allowing for the absorption of
ambiguities into an on shell perturbative renormalization group,
of the  following type: there is an $SL2 \mathbb{C}$ invariant
representative $\calw$ of the Wick algebra as a subspace of
$\calp$, the algebra of local polynomials such that the Ward
identity reads
\begin{equation}\label{eq2bis}
    \cald^{x_1} \ T \ w_1(\varphi)(x_1) \ldots w_n(\varphi) (x_n)= T \
    \cald^{x_1}
    w_1(\varphi) {(x_1)} \ldots w_n (\varphi)(x_n)
\end{equation}
for $w_i(\varphi) \in \calw$, for all differential operators
$\cald^x$ such that $\cald \ w_1(\varphi) \subset \calw$ and such
that $T$ vanishes on shell whenever any of its arguments vanishes
by virtue of the free field equations of motion, so that such
$T$'s define some $\widehat{T}$.

 Remark: \ref{section32} and \ref{section33} are not
related to symmetries.

\subsection{Exact or softly broken global internal
symmetries\cite{15}(abc)\label{section34}}

The corresponding Ward identities can always be fulfilled for
semisimple internal symmetry groups, again by $H^1 (Lie \ G, F)
=0$ for semisimple $Lie G$. If $U(1)$ factors occur, besides the
Wess Zumino consistency conditions, renormalization group
considerations are necessary.

\subsection{Current algebra\cite{15}\label{section35}}

This is a sequel to \ref{section34}.

The Wess  Zumino consistency condition was solved for an arbitrary
$G$, for renormalizable theories as early as 1975\cite{15}(e).
That was a central part in the renormalization of quantized gauge
theories. It took quite some time to release the power counting
restrictions implied by renormalizability\cite{16}.

\subsection{Gauge theories\label{section36}}

This is a sequel to \ref{section35}!

The Slavnov Taylor identity buries the gauge Lie algebra structure
under an exotic layer of auxiliary fields whose role is to perform
a sequence of quotients whereas locality is saved. There results a
differential algebra whose relevant cohomology ($H^0$ and $H^1$)
reduces to the corresponding part of the gauge Lie algebra
cohomology.

The Faddeev Popov ghost field appears as the generator of the
gauge Lie algebra cohomology.

The Faddeev Popov antighost together with a Nakanishi Lautrup
multiplier field are associated with gauge fixing. Failing to
include the latter results into a differential up to the antighost
equation of motion, which is rather inconvenient\cite{19}.

Finally (this should actually come first!) the antifields are
needed to express the fact that gauge invariance only holds up to
the equations of motion.

The exotism of this collection of fields which allows to recover
gauge invariance after its breaking through gauge fixing allows
for the systematic use of the Wess Zumino consistency condition,
pertaining to the Slavnov Taylor identity which is better
(although not completely) understood now than when it first
appeared.

\subsection{"Field dependent" Lie algebras\label{section37}}

All the preceding examples were associated with clearly defined
algebraic structures encoded into a differential algebra. It is
often not obvious how to reach such a clear situation. Here are a
few more examples.

\subsubsection{Gravity\label{section37a}}

The variables are taken as a vierbein $e$ and a spin correction
$\omega_{M_4}$ both locally defined one forms on space time $M_4$
and acted upon both by vector fields $\xi$ on $M_4$ and $SL2
\mathbb{C}$ gauge transformations $\Omega$.

The structure equations for "parallel transport" read
\begin{eqnarray}
\call (\xi, \Omega) e &=& i_\xi T + \Omega e+D_\omega(i_\xi e)
\label{eq3}\\
\call (\xi, \Omega) w &=& D_\omega \Omega + i_\xi R \label{eq4}
\end{eqnarray}
where
\begin{eqnarray}
T &=& de +\omega e, \ \ R=d\omega + \frac{1}{2}  [\omega,\omega] \label{eq6} \\
D_\omega &=& d+t(\omega) \label{eq6}
\end{eqnarray}
where $t$ is the relevant representation of Lie$(SL2 \mathbb{C})$
and
\begin{equation}\label{eq7}
    i_\xi = \mbox{contraction of form with vector field.} \ \xi
    \label{eq7}
\end{equation}

The commutation relations are
\begin{eqnarray}
\left[ \call (\Omega), \call (\Omega') \right] &=& \call \left(
\left[ \Omega,\Omega' \right] \right)
\label{eq8} \\
\left[ \call(\Omega), \call(\xi) \right] &=&0 \label{eq9} \\
\left[ \call (\xi), \call(\xi') \right] &=& \call \left( \left[
\xi,\xi' \right] \right) + \call \left( i_\xi i_{\xi'} R \right)
\label{eq10}
\end{eqnarray}
where
\begin{equation}\label{eq11}
    [\xi,\xi']= \mbox{commutator of vector fields} \ \xi,\xi'.
\end{equation}

The second commutation relation (eq.\ref{eq9} might lead one to
think that\cite{18} at the infinitesimal level diffeomorphisms
commute with $SL2 \mathbb{C}$ gauge transformations, which is of
course not right: the $\call$'s do not generate a Lie algebra
because of the field dependent curvature term in the last
commutator [eq.\ref{eq10}].

What are then the Wess Zumino consistency conditions?

The answer is the following: if one lifts $\xi$ along the
horizontal planes of a background connection
$\stackrel{0}{\omega}$ one produces a Lie algebra
$\stackrel{0}{\call}$.
\begin{eqnarray}
\left[ \stackrel{0}{\call} (\Omega) , \stackrel{0}{\call}
(\Omega') \right] &=&
\stackrel{0}{\call} \left( [\Omega,\Omega'] \right) \label{eq11} \\
\left[ \call^0 (\xi) , \call^0 (\Omega) \right] &=& \call^0 \left(
\stackrel{0}{\call}_{(\xi)} \Omega \right) = \stackrel{0}{\call}
\left( i_\xi
D_{\stackrel{0}{\omega}} \Omega \right) \label{eq12} \\
\left[ \call^0 (\xi) , \call^0 (\xi') \right] &=&
\stackrel{0}{\call} \left(
 [\xi,\xi'] \right) - \dot{\call} \left( i_\xi i_\xi
\stackrel{0}{R} \right) \label{eq13}
\end{eqnarray}
where
\begin{eqnarray}
D_{\stackrel{0}{\omega}} &=& d + t(\stackrel{0}{\omega}) \label{eq14} \\
\stackrel{0}{R} &=& d\stackrel{0}{\omega} + \frac{1}{2}
[\stackrel{0}{\omega} , \stackrel{0}{\omega}]. \label{eq15}
\end{eqnarray}
$\stackrel{0}{\call}$ seems to depend on $\stackrel{0}{\omega}$.
In fact, it does not: there is a Lie algebra $\eps$, independent
of $\stackrel{0}{\omega}$: one goes from $\stackrel{0}{\call}$ to
$\stackrel{1}{\call}$ by the change of generators
\begin{equation}\label{eq16}
\stackrel{0}{\call} (\xi) = \stackrel{1}{\call} (\xi) + \call
(i_\xi (\stackrel{0}{\omega} - \stackrel{1}{\omega}))
\end{equation}
\begin{equation}\label{eq17}
    \left( \stackrel{0}{\call} (\Omega) = \stackrel{1}{\call} (\Omega) = \call (\Omega)
    \right).
\end{equation}

This, one knew beforehand: all objects sit on a bundle $P$. At the
Lie algebra level, the automorphisms of the bundle are given by an
extension of the diffeomorphisms of the base by gauge
transformations
\begin{equation}\label{eq18}
    0 \rightarrow Lie \  \Omega \rightarrow Lie \ Aut \ P
  \stackrel{\stackrel{0}{\omega}}{\rightleftharpoons} \ \ Vect \ M \rightarrow 0
\end{equation}

A choice of  $\stackrel{0}{\omega}$ defines (non canonically) a
lift of Vect $M$ into $\eps$. This extension is neither central
nor abelian.

Parallel transport is best defined within the cohomology algebra
of $\eps$ with values functionals of the vierbein and the spin
connection by a change of generators
\begin{equation}\label{eq19}
    \Omega \rightarrow \Omega +i_\xi (\omega-\stackrel{0}{\omega})
\end{equation}
which is quite a licit operation whereas
\begin{equation}\label{eq20}
    \call^0 (\xi) \rightarrow \stackrel{0}{\call} (\xi) + \call
    (i_\xi(\omega-\stackrel{0}{\omega}))
\end{equation}
is quite strange since it mixes up parameters labelling the
transformations with the field variables they act on\cite{19}.

\subsubsection{Anomalies in the on shell operator
formalism\label{section37b}}

Whereas the locality properties of the off shell  formalism offer
a good framework for the discussion of anomalies through the WZCC,
it is much less so in the on shell formalism, from which there
results algebraically hybrid situations which have been
investigated in the case of free strings\cite{20} (M. Kato, K.
Ogawa, 1983, rigorized by I.B. Frenkel, H. Garland, G.J.
Zuckerman, 1986)\cite{20}, where the result is exact, and by L.D.
Faddeev\cite{21} (1984) in the gauge case, at a semi-classical
level.

In both cases, the anomaly manifests itself either at the gauge
Lie algebra level as a field dependent cocycle which characterizes
an extension of the gauge Lie algebra (in the string case, it is
field dependent because the Faddeev Popov ghost is quantized), or,
equivalently, as a lack of nilpotency of a BRST charge (both the
definition of the charge and of its square require ad hoc
renormalizations).

The situation is hybrid because the anomaly is a cocycle for the
initial gauge Lie algebra.

As expected, the off shell version of this phenomenon is
algebraically subject to the standard WZCC: it is simply the study
of anomalies for the BRS current which can be shown to be an
algebraic extension of the usual anomaly of the Slavnov identity
alluded to in \ref{section36}\cite{22} (L. Baulieu, B. Grossmann,
R. Stora, 1986).

Besides these algebraic formalities, it is worth mentioning that
this class of examples can be described as an infinite dimensional
version of \ref{section37a}: the base space is configuration space
of classical fields acted upon by a gauge group, and above it a
complex line bundle ("DET") (whose sections are functionals of the
fields), with structure group $U(1)$ (phases). The anomaly is
precisely of the form of the extra curvature term in
Eq.\ref{eq10}, whose locality in the field variables is related to
general properties of index theorems\cite{23}.

\subsection{A General Receptacle : BV\label{section38}}

From the perturbative renormalization of gauge theories based on
locality and power counting, we have learnt that besides gauge and
matter fields we need:
\begin{itemize}
    \item $\phi\pi$ ghosts, associated with infinitesimal gauge
    transformations,
    \item $\phi\pi$ antighost fields and Nakanishi Lautrup
    Lagrange multipliers, associated with gauge fixing
    \item "antifields", coupled to the non linear gauge
    transformations with parameter $\phi\pi$ ghost, and allowing
    their deformation -renormalization.
\end{itemize}

It was remarked by J. Zinn-Justin (1974) that writing the
corresponding Ward identity (in this case called the Slavnov
Taylor identity) for the vertex functional leads to its
interpretation in terms of a graded symplectic -resp. contact-
structure already met in mathematics in the context of Hochschild
cohomology under the denomination "Gerstenhaber bracket", and
further studied by I.V. Batalin, G.A. Vilkovsky (1981)\cite{24}.

The role of the antifields is to describe, with due respect of
locality, symmetries which hold "modulo the equations of motion"
("on shell")\cite{24}\cite{16}. This formalism has been
subsequently applied to a variety of theories of the gauge type,
including such exotic classes as supergravities, A.
Zamolodchikov's $W$ algebras, M. Konsevitch's quantization of
Poisson structures, to quote only a few.

Needless to say, the WZCC have accordingly propagated into the
study of the corresponding cohomologies.

\section{Concluding Remarks \label{section4}}

The innocent looking Wess Zumino consistency conditions have
turned out to be an essential tool, at least -but not only- within
the framework of renormalized perturbation theory, in the
discussion of the physical content of field theory models founded
on locality. There, restrictions of the ambiguities inherent to
the singularity of local interactions experimentally proceed via
the imposition of a variety of Ward identities and a discussion of
their potential anomalies which, when present, precludes the use
of any general intelligent regularization procedure. Various
standard algebraic set ups have occured in field theory, beyond
Lie algebras associated with symmetries, whose mathematics has
been known for more than half a century, now, and together with
them, a zoo of cohomologies for which the Wess Zumino consistency
conditions are nothing else than cocycle conditions. Many more are
expected to appear for the simple reason that in the sort of
theoretical physics under practice, the enforcement of general
principles -e.g. locality- delivers physical predictions as
equivalence classes -resp. quotients- of objects belonging to a
much larger class than that of the observables.

So, long life to WZCC!


\begin{thebibliography}{99}
\bibitem{5}
This body of knowledge is globally referred to as the BPHZ
framework.\\
Locality is most apparent in refs. \cite{6}, \cite{7}, \cite{8}
from which the past history can be recovered, but the
combinatorics is most developed in $Z$ (ref.\cite{9}), summarized
in \cite{10}.
\bibitem{18} There is an amusing story going along with this,
whose actors may remember.
\bibitem{19}
These are examples where difficulties to write down WZCC signal an
improper -incomplete- algebraic setup. The following examples also
bring some water to this mill. It is tempting to require as a
principle that the Wess Zumino algebraic consistency be fulfilled
in further constructions of that sort.
\end{thebibliography}
\end{document}